# THE PLASMA-SOLID EQUILIBRIUM IN A DEBYE PLASMA

V.^elebonovi}

Institute of Physics,Pregrevica 118,11080 Zemun,Yugoslavia

## 1. INTRODUCTION

Observational data collected by various astronomical methods in the last two centuries point to the fact that matter in the universe exists not only in the form of celestial objects ( from asteroids to galaxies),but also in the form of various kinds of bright and dark clouds. Investigations of these clouds performed in the last century have unambiguously shown that they are the places where stars and planets are being born. For a recent review of the interstellar medium see, for example [1] .

As the universe contains different kinds of plasmas, but also solid objects, it logically follows that there must exist a region in which a phase transition between the two regimes takes place. For obvious reasons, this transition is called the "plasma-solid" transition (PS transition),and the aim of this communication is to contribute to its study.

In previous work on this subject [2], the Debye temperature of a solid in equilibrium with the Fermi gas surrounding it was calculated. The present work represents a step further, in the sense that the plasma was modeled in a more realistic way, within the Debye-Hückel (DH) theory.

## 2. CALCULATIONS

The energy of a DH plasma is given by [3]

$$E = \frac{3}{2}\sum_j V n_{j0} k_B T - \frac{V k_B T}{8\pi l_D^3} \qquad (1)$$

In this equation,the symbol $n_{j0}$ denotes the number density of of particles of type $j$ ,and $l_D$ is the Debye length defined by [3]

$$l_D^{-2} = \frac{4\pi e^2}{k_B T}\sum_j n_{j0} z_j^2 \qquad (2)$$

It follows after some algebra that the energy per particle of a DH plasma can be expressed as follows:

$$E_{P,D} = \frac{E}{N} = \frac{3}{2}k_B T(1 - \frac{1}{12\pi n_0 l_D^3}) \qquad (3)$$

It can be shown [2] (and references given there) that the energy per particle of a solid is, in Debye's model, given by

$$E_{P,S} \cong 3k_B T[1 + \frac{1}{20}(\frac{\theta}{T})^2 - \frac{1}{1680}(\frac{\theta}{T})^4] \qquad (4)$$

where $\theta$ denotes the Debye temperature.
Detailed considerations ([2],[4]) have shown that the criterion for the equilibrium of a solid with a plasma in which it is embedded can be formulated as:

$$f = \frac{E_{PS}}{E_{PD}} = 1 \qquad (5)$$

This ration can easily be formed from eqs.(3) and (4). Solving the resulting expression for the Debye temperature, it follows:

$$\theta^2 = 42T^2 \pm \frac{T^2}{l_D^{3/2}}\sqrt{\frac{14}{\pi n}(5 + 186n\pi l_D^3)} \qquad (6)$$

which is the expression for the Debye temperature of a solid in equilibrium with a DH plasma.

## 3. DISCUSSION

The aim of this brief calculation is to contribute to knowledge of the PS transition in various kinds of astrophysical plasmas. Therefore, the first step in discussing eq.(6) has to be the verification of the applicability of the DH model to space plasma.

A general condition for the applicability of the DH theory is [3]

$$l_D \geq r_0 \qquad (7)$$

where $r_0$ denotes the average distance between the particles. This implies [3]

$$n_0 \leq \left(\frac{k_B T}{4\pi e^2 z^2}\right)^3 \approx 10^5 \left(\frac{T}{z^2}\right)^3 cm^{-3} \qquad (8)$$

It is known from observation that in cold interstellar clouds in our galaxy temperatures are of the order of $10 \leq T[K] \leq 20$ and the number densities are in the range $10^4 \leq n[cm^{-3}] \leq 10^5$. Inserting these data to eq.(8) shows that the DH theory is applicable for $n \prec 10^8 cm^{-3}$, which implies that it can be used in studies of interstellar clouds.

The physical motive for the calculations dicussed in this paper is clear: to determine the Debye temperature of a solid in equilibrium with a DH plasma. As the Debye temperature is a unique parameter of every solid, this could be a useful result in interpretation of observational data on interstellar clouds (such as [5] ). Inserting the values of various constants in eq. (6) finally gives:

$$\boldsymbol{q}^2 = 42T^2 \pm 0.01436 T^{1/2} n Z^3 \qquad (9)$$

The Debye temperature is real, and it follows from eq.(9) that the solution with the minus sign is applicable for

$$T \geq 0.00488956 n^{2/3} Z^2 \qquad (10)$$

For example, inserting $n = 10^5; Z = 1$ in eq.(10) gives $T \geq 10.5 K$, which is of the order of magnitude of the temperature measured in the coldest molecular clouds in our galaxy. This implies that both solutions for the Debye temperature given by eq.(9) can be used in work on real interstellar clouds. Inserting $T = 20K; n = 10^5 cm^{-3}; Z = 2$ in eq.(9) with the positive sign gives $\boldsymbol{q} \approx 261K$, which is a physically reasonable value. Increasing the temperature to 45 K, leads to the value of the Debye temperature of $\boldsymbol{q} \approx 402K$. This last value is close to the experimental value for the chemical element Si, and it could be interpreted as meaning that Si can exist in equilibrium with a DH plasma having these values of the temperature and particle number density. Furthermore, silicon grains have been detected in various interstellar clouds ( [2] and references given there). Varying the values of the temperature and number density which enter eq.(9), one could obtain the value of the Debye temperature which corresponds to any given chemical element and/or compound. This could be taken to imply that if this solid exists in a plasma specified by a given pair $(T, n)$ it is in equilibrium with the plasma. Turning things to pure astrophysics, this could have useful implications for studies of planetary formation .

On the side of pure physics, at least two open points will be discussed in future work on this subject: the density encountered in the calculations discussed in this contribution is actually a sum of the densities of various particle species. The obvious question is how does the mixture of several kinds of charged particles influence the Debye temperature? Another point is to determine if there are there any physical limitations for the application of the two solutions given in eq.(9). The condition in eq.(10) is just a mathematical consequence of eq.(9). More details will be published elsewhere.